\RequirePackage{lineno}
\documentclass[aps,prl,twocolumn,superscriptaddress,showpacs,preprintnumbers,amsmath,amssymb]{revtex4-2}
\usepackage[utf8]{inputenc}

\usepackage[dvipsnames]{xcolor}
\usepackage[normalem]{ulem}
\usepackage{comment}
\usepackage{overpic}
\usepackage{multirow}
\usepackage{booktabs} 

\graphicspath{{./figures/}}
\usepackage[colorlinks,linkcolor=red,anchorcolor=blue,citecolor=red,breaklinks=true]{hyperref}
\usepackage{cleveref}
\newsavebox{\tablebox}
\usepackage{orcidlink}

\input{alias.tex}

\def\to {\!\rightarrow\!}
\def\KS {K^0_S}
\def\DpCS {{D^{+}\to K^{+}K^{-}\pi^{+}\pi^{0}}}
\def\DpDCS {{D^{+}\to K^{+}\pi^{-}\pi^{+}\pi^{0}}}
\def\DpCF {{D^{+}\to K^{-}\pi^{+}\pi^{+}\pi^{0}}}
\def\DsCS {{D_s^{+}\to K^{+}\pi^{-}\pi^{+}\pi^{0}}}
\def\DsCF {{D_s^{+}\to K^{+}K^{-}\pi^{+}\pi^{0}}}
\def\DzKK {{D^0\to K^{+}K^{-}}}
\def\Dzpp {{D^0\to\pi^{+}\pi^{-}}}
\def\DpDsp {D^+_{(s)}}

\begin{document}

\title{Measurement of the branching fractions for Cabibbo-suppressed decays ${D^+\to\Kp K^{-}\pip\piz}$ and ${D_{(s)}^+\to\Kp\pi^{-}\pip\piz}$ at Belle} 

\noaffiliation

  \author{L.~K.~Li\,\orcidlink{0000-0002-7366-1307}} 
  \author{A.~J.~Schwartz\,\orcidlink{0000-0002-7310-1983}} 
  \author{K.~Kinoshita\,\orcidlink{0000-0001-7175-4182}} 
  \author{E.~Won\,\orcidlink{0000-0002-4245-7442}} 

  \author{H.~Aihara\,\orcidlink{0000-0002-1907-5964}} 
  \author{S.~Al~Said\,\orcidlink{0000-0002-4895-3869}} 
  \author{D.~M.~Asner\,\orcidlink{0000-0002-1586-5790}} 
  \author{H.~Atmacan\,\orcidlink{0000-0003-2435-501X}} 
  \author{V.~Aulchenko\,\orcidlink{0000-0002-5394-4406}} 
  \author{T.~Aushev\,\orcidlink{0000-0002-6347-7055}} 
% ...
\collaboration{The Belle Collaboration}


\begin{abstract}
We present measurements of the branching fractions for the singly Cabibbo-suppressed decays $\DpCS$ and $\DsCS$, 
and the doubly Cabibbo-suppressed decay $\DpDCS$, based on 980~$\rm fb^{-1}$ of data 
recorded by the Belle experiment at the KEKB $e^+e^-$ collider. 
We measure these modes relative to the Cabibbo-favored modes 
$\DpCF$ and $\DsCF$. Our results for the ratios of branching fractions are 
${\mathcal{B}(\DpCS)/\mathcal{B}(\DpCF)} = {(11.32 \pm 0.13 \pm 0.26)\%}$,
${\mathcal{B}(\DpDCS)/\mathcal{B}(\DpCF)} = {(1.68 \pm 0.11 \pm 0.03)\%}$, and 
${\mathcal{B}(\DsCS)/\mathcal{B}(\DsCF)} = {(17.13 \pm 0.62 \pm 0.51)\%}$,
where the uncertainties are statistical and systematic, respectively. 
The second value corresponds to
${(5.83\pm0.42)\times\tan^4\theta_C}$,
where $\theta_C$ is the Cabibbo angle; 
this value is larger than other measured ratios 
of branching fractions for a doubly Cabibbo-suppressed 
charm decay to a Cabibbo-favored decay.
Multiplying these results by world average values for
$\mathcal{B}(\DpCF)$ and $\mathcal{B}(\DsCF)$ yields 
$\mathcal{B}(\DpCS)= {(7.08\pm 0.08\pm 0.16\pm 0.20)\times10^{-3}}$, 
$\mathcal{B}(\DpDCS)= {(1.05\pm 0.07\pm 0.02\pm 0.03)\times10^{-3}}$, and 
$\mathcal{B}(\DsCS) = {(9.44\pm 0.34\pm 0.28\pm 0.32)\times10^{-3}}$, 
where the third uncertainty is due to the branching fraction of the
normalization mode. The first two results are consistent with, but more
precise than, the current world averages. The last result is the first measurement of this branching fraction.

\end{abstract}



\maketitle

\section{Introduction}

Cabibbo-suppressed~(CS) hadronic decays of charm mesons provide a 
powerful means to search for new physics~\cite{bib:CPV1}. Because 
such decays are suppressed in the Standard Model, their decay rates 
are especially sensitive to small new-physics contributions to the amplitudes. 
Thus, it is important to measure such decays with high precision. It 
is notable that the only observation of $\CP$ violation in charm 
decays, possibly arising from new physics~\cite{bib:CPV1},
was made with the singly Cabibbo-suppressed~(SCS) decays 
$\DzKK$ and $\Dzpp$~\cite{bib:CPVobservation}. Experimentally, 
CS decays can be challenging to measure, as they typically have 
higher background levels than those for Cabibbo-favored~(CF) decays. 

In this paper, we present measurements of the branching fractions for 
the SCS decays $\DpCS$ and $\DsCS$, and the 
doubly Cabibbo-suppressed~(DCS) decay $\DpDCS$. 
Throughout this paper, charge-conjugate modes are implicitly included.
The branching fractions are measured relative to those for the 
well-measured CF modes $\DpCF$ and $\DsCF$. 
The branching fraction of a DCS decay relative to 
its CF counterpart is expected to be approximately
$\tan^4\theta_C\!=\!0.29\%$~\cite{bib:PRD81.074021},
where $\theta_C$ is the Cabibbo angle.
The SCS decay $\DpCS$, and the DCS decay $\DpDCS$, were recently observed 
by the BESIII experiment~\cite{bib:PRD102d052006,bib:PRL125D141802,bib:arxiv2105D14310}.
The decay $\DsCS$ has not yet been observed.  
The average of the absolute branching fractions measured at BESIII~\cite{bib:PRL125D141802,bib:arxiv2105D14310} is $\mathcal{B}(\DpDCS)\!=\!{(1.18\pm0.07)\times10^{-3}}$;
this gives a ratio of branching fractions
${\mathcal{B}(\DpDCS)/\mathcal{B}(\DpCF)}\!=\!{(1.89\pm0.12)\%}$, 
which corresponds to $(6.56\pm 0.42)\times\tan^4\theta_C$.
This value is larger than other measured ratios 
of DCS to CF branching fractions, 
which are in the range
${(0.7\text{--}1.8)\times\tan^4\theta_C}$~\cite{bib:PDG2021}.
To investigate this further, 
we use the full Belle data set to measure these decay
modes  with high precision.

\section{Detector and data set}
\label{sec:data}

Our analysis uses the full data set of the Belle experiment, 
which corresponds to an integrated luminosity of 980~$\invfb$ 
collected at or near the $\Upsilon(nS)$ ($n=1,\,2,\,3,\,4,\,5$) resonances.
The Belle experiment ran
at the KEKB asymmetric-energy $\epem$ collider~\cite{bib:KEKB,bib:Bfactories}.
The Belle detector is a large-solid-angle magnetic spectrometer 
consisting of a silicon vertex detector~(SVD), a $50$-layer central drift chamber~(CDC), 
an array of aerogel threshold Cherenkov counters~(ACC), a barrel-like arrangement of 
time-of-flight scintillation counters~(TOF), and an electromagnetic calorimeter~(ECL) 
comprising CsI(Tl) crystals located inside a superconducting solenoid coil providing 
a $1.5$~T magnetic field. An iron flux-return located outside the coil is instrumented 
to detect $K_L^0$ mesons and to identify muons~(KLM).
A detailed description of the detector is given in Ref.~\cite{bib:Bfactories,bib:BelleDetector}. 

We use Monte Carlo~(MC) simulated events to optimize selection criteria, study 
sources of background, and calculate selection efficiencies. 
Signal MC events are generated using {\sc EvtGen}~\cite{bib:evtgen} 
and propagated through a detector simulation based on {\sc Geant3}~\cite{bib:geant3}. 
Final-state radiation from charged particles is simulated using {\sc Photos}~\cite{bib:PHOTOS}. 
Four-body decays are generated to decay uniformly in phase space 
without intermediate resonances.
An MC sample of generic $e^+e^-$ collisions corresponding to the same 
integrated luminosity as the data sample is used to develop selection 
criteria.

\section{Event selection}
\label{sec:selection}
To ensure that tracks are well reconstructed, each final-state charged particle is required to have at least two SVD hits in each of the longitudinal and azimuthal measuring coordinates.
Charged particles are identified by calculating 
likelihoods $\mathcal{L}_i$ for specific particle hypotheses, 
where $i=\pi,\,K,\,p,\,\mu,\,e$. These likelihoods are based on 
information from various detectors:
photon yield in the ACC, $dE/dx$ information from the CDC, 
time-of-flight information from the TOF,
energy in the ECL, and hits in the KLM~\cite{bib:PID,bib:NIMA485D490,bib:NIMA491d69}.
Tracks with $\mathcal{L}_K/(\mathcal{L}_K+\mathcal{L}_{\pi}) > 0.6$ are 
identified as kaon candidates; otherwise, tracks are considered pion candidates.
Kaon candidates must also satisfy $\mathcal{L}_p/(\mathcal{L}_p+\mathcal{L}_{K}) < 0.95$.
Tracks that satisfy 
$\mathcal{L}_e/(\mathcal{L}_e+\mathcal{L}_{\rm hadron})>0.95$
or 
$\mathcal{L}_\mu/(\mathcal{L}_\mu+\mathcal{L}_{\pi}+\mathcal{L}_{K})>0.95$
are rejected,
where $\mathcal{L}_e$, $\mathcal{L}_{\rm hadron}$, and $\mathcal{L}_\mu$
are determined mainly using information from the ECL and KLM
detectors~\cite{bib:NIMA485D490,bib:NIMA491d69}.
These requirements have an efficiency of about $90\%$ for kaons and $95\%$ for pions.

Photon candidates are identified from energy clusters in the ECL 
that are not associated with any charged track. The photon energy
is required to be greater than 50~MeV in the barrel region (covering the
polar angle $32^\circ < \theta < 129^\circ$), and greater than 100~MeV 
in the endcap region (covering $12^\circ < \theta < 31^\circ$ or $132^\circ < \theta < 157^\circ$). 
The ratio of the energy deposited in the 3$\times$3 array of crystals centered on the 
crystal with the highest energy, to the energy deposited in the corresponding 
5$\times$5 array of crystals, is required to be greater than~0.80.
Candidate ${\piz\to\gamma\gamma}$ decays are reconstructed from photon pairs having an
invariant mass satisfying $115~{\rm MeV}/c^2 < M(\gamma\gamma) < 150~{\rm MeV}/c^2$; 
this region corresponds to about $3\sigma$ in $M(\gamma\gamma)$ resolution. 

A $\DpDsp$ candidate is reconstructed by combining $\Km\Kp\pip$ or $K^{\pm}\pi^{\mp}\pip$ 
track combinations with a $\piz$ candidate. 
A vertex fit is performed for the three charged tracks and its fit quality is defined as $\chi^2_{\rm vtx}$. The coordinates of the fitted vertex are assigned as the $\DpDsp$ decay vertex position.

Final selection criteria are determined by maximizing a 
figure-of-merit (${\rm FOM}$), which is defined as either 
$S/\sqrt{S+B}$ for $\DpCS$ and $\DpDCS$ or
$S/\sqrt{B}$ for $\DsCS$, 
where $S$ and $B$ are the numbers of signal 
and background events, respectively, expected in a 
region $-30~{\rm MeV}/c^2  < M(D)-m_D < 20~{\rm MeV}/c^2$.
In this expression, 
$M(D)$ is the invariant mass of a reconstructed $\Dp$ or $\Dsp$ 
candidate, and $m_D$ is the known $\Dp$ or $\Dsp$ mass~\cite{bib:PDG2021}. 
This region corresponds to about $2.5\sigma$ in the $M(D)$ resolution.
The FOM for $\DsCS$ is different because the branching fraction for 
this mode has not yet been measured.

Pairs of $\gamma$ candidates are subjected to a fit in which the $\gamma$'s are constrained 
to originate from the $\DpDsp$ decay vertex, 
and their invariant mass is constrained to the nominal $\piz$ mass~\cite{bib:PDG2021}.
The resulting fit 
quality ($\chi^2_{\piz}$) is required to satisfy $\chi^2_{\piz}<8$.
To improve the momentum resolution of the $\piz$, the $\gamma$ energies are 
updated from this fit; the resulting $\piz$ momentum is required to be greater than 0.40~GeV/$c$.
We veto $\DpDsp\to\Kp\pim\pip\piz$ candidates satisfying $|M(\pip\pim)-m^{}_{\KS}|<10~{\rm MeV}/c^2$, 
where $m^{}_{\KS}$ is the nominal
$\KS$ mass~\cite{bib:PDG2021}, to suppress 
peaking backgrounds such as $\DpDsp\to\Kp\KS\piz$. 
This region corresponds to about $3\sigma$ in mass resolution.

The $\DpDsp$ production vertex is determined by fitting the 
$\DpDsp$ trajectory to the $e^+e^-$ interaction point (IP), 
which is determined from the beam profiles. 
This vertex fit quality is defined as $\chi^2_{\rm IP}$. 
The sum of vertex fit qualities $\chi^2_{\rm vtx}+\chi^2_{\rm IP}$ is 
required to be less than 14 for $\DpCS$ decays, and less than 10 
for the other signal modes. 
This requirement has a signal efficiency of 
80\%--82\% while rejecting 60\%--80\% of background.

The dominant source of background is random combinations 
of particles produced in $e^+e^-\to c\bar{c}$ events or 
in $B$ decays. To suppress this background, 
the momentum of the $\Dp$ or $\Dsp$ candidate
in the $\epem$ center-of-mass frame is required 
to be greater than 2.5~GeV/$c$ or 2.9~GeV/$c$,
respectively.
To further suppress backgrounds, we calculate the significance 
of the $\DpDsp$ decay length $L/\sigma_L$, where 
$L$ is the projection of the vector running from 
the production vertex to the $\DpDsp$ decay vertex
onto the momentum direction. The corresponding uncertainty 
$\sigma_L$ is calculated by propagating uncertainties in 
the vertices and the $\DpDsp$ momentum, including their correlations. 
We subsequently require 
$L/\sigma_L\!>\!4.0$ for $\DpCS$, 
$L/\sigma_L\!>\!9.0$ for $\DpDCS$, and 
$L/\sigma_L\!>\!2.5$ for $\DsCS$.
The resulting signal efficiencies are 
58\%--77\%, while more than 93\%--99.8\% of background is rejected. 

The CF normalization modes $\DpCF$ and $\DsCF$ 
are selected with the same criteria as those used to select the 
signal modes, to minimize systematic uncertainties. For both 
signal and normalization modes, we retain events that satisfy 
$-70~{\rm MeV}/c^2 < M(D)-m_D < 60~{\rm MeV}/c^2$.

After applying all selection criteria, about 10\% of 
events for $D^+$ decay modes and 15\% of events for $D_s^+$ decay modes
have multiple signal candidates. For these events, the average 
multiplicity is about 2.2 candidates for each channel.
We select a single candidate by choosing the one with the 
smallest value of the sum $\chi^2_{\piz} + \chi_{\rm vtx}^2 + \chi_{\rm IP}^2$.
Based on MC simulation, this criterion selects the correct signal candidate 
68\% of the time.

There are backgrounds from $D^{*+}$ decays in which the final state particles 
are the same as those for the signal or normalization modes. These are as follows:
\begin{itemize}
\item for ${D^+\to\Km h^+\pip\piz}$ decays, where $h^+\!=\!K^+$ or $\pi^+$, there 
is background from ${D^{*+}\to\Dz\pi^+}$, ${\Dz\to\Km h^+\piz(\piz)}$.To reject this 
background, we require $M({\Km h^+\pip\piz})-M({\Km h^+\piz})-m_{\pip}\!>\!20~{\rm MeV}/c^2$.
\item for $\DsCF$, there is background from 
${D^{*+}\to D^0\pi^+}$, ${D^0\to\Km\Kp\pi^0}$, and from 
${D^{*+}\to D^+\pi^0}$, ${D^+\to\Km\Kp\pi^+}$.  
To reject these, we require 
$M({\Km\Kp\pip\piz})-M({\Km\Kp\pi^+})-m_{\pi^0}\!>\!10~{\rm MeV}/c^2$, and also
$M(\Km\Kp\pip\piz)-M(\Km\Kp\pi^0)-m_{\pi^+}\!>\!10~{\rm MeV}/c^2$.
\item for ${\DpDsp\to\Kp\pim\pip\piz}$, there is background from
${D^{*-}\to\Dzb\pi^{-}}$, ${\Dzb\to\Kp\pim\piz}$, with the $\pim$ replaced by a random $\pip$. To suppress this background, we require $M(\Kp\pim\pip\piz)-M(\Kp\pip\piz)-m_{\pim}\!>\!40~{\rm MeV}/c^2$.
\end{itemize}
\noindent These requirements 
reject only 1\%--3\% of signal decays
but reduce $D^{*+}$ backgrounds to a negligible level.  

\section{Yield extraction}
\label{sec:yields}

We determine signal yields by performing an extended unbinned 
maximum-likelihood fit to the $M(D)$ distributions. 
The probability density function~(PDF) describing signal decays is taken to be
the sum of a Crystal Ball function~\cite{bib:CB} and three asymmetric Gaussians~(AG), which are Gaussian functions with different widths on the left- and right-hand sides of the peak position. This position is denoted by the parameter $\mu$, and all four functions are required to have a common value of $\mu$. 
An additional term ($\mathcal{P}_{\rm FSR}$) is 
included to describe signal decays with final-state radiation~(FSR). 
For this term, the sum of a CB function and a Gaussian is used; the 
parameters of $\mathcal{P}_{\rm FSR}$ and its ratio to the total signal 
yield ($f_{\rm FSR}$) are fixed to MC values.
The overall PDF is 
\begin{eqnarray}
\mathcal{P}_{\rm sig} & = & 
(1-f_{\rm FSR}) \bigg[ f_{3} \Big[ f_{2} \big[ f_{1} \cdot {\rm AG}(\mu, \sigma_1, \delta_1) \nonumber \\
& & \hskip1.10in +\ (1-f_{1})\cdot {\rm AG}(\mu, \sigma_{2}, \delta_{2})\big] \nonumber \\
& & \hskip0.90in +\ (1 - f_{2})\cdot {\rm AG}(\mu, \sigma_3, \delta_3) \Big]  \nonumber \\
& & \hskip0.90in +\ (1-f_{3})\cdot {\rm CB}(\mu, \sigma_{4}, \alpha_{\rm cb}, n_{\rm cb}) \bigg]  \nonumber \\
& &  \hskip0.9in +\ f_{\rm FSR}\cdot\mathcal{P}_{\rm FSR}\,, \label{eqn:sigpdf}
\end{eqnarray}
where $\sigma_{i+1}\!=\!r_i \sigma_{i}$ ($i\!=\!1,\,2,\,3$) with a scaling factor $r_i$, 
and the left-side~($L$) and right-side~($R$) widths of 
the asymmetric Gaussians are specified by the parameter $\delta_i$: 
$\sigma_{i}^{L,R}=\sigma_i(1\pm\delta_i)$. 
The parameters $\mu$ and $\sigma_1$ are free to vary, which allows for a difference in resolution between data and 
MC simulation; all other parameters are fixed to MC values.
The mode $\DpDCS$ is fitted simultaneously with the normalization 
mode $\DpCF$, as both modes share signal shape parameters.

The background shapes are described by second-order Chebyshev polynomials for 
$\DpDCS$ and $\DsCS$, and a third-order Chebyshev polynomial for $\DpCS$.
All parameters of these shapes are free to vary. 

\begin{figure}[!htbp]
  \begin{centering}%
   \begin{overpic}[width=0.45\textwidth]{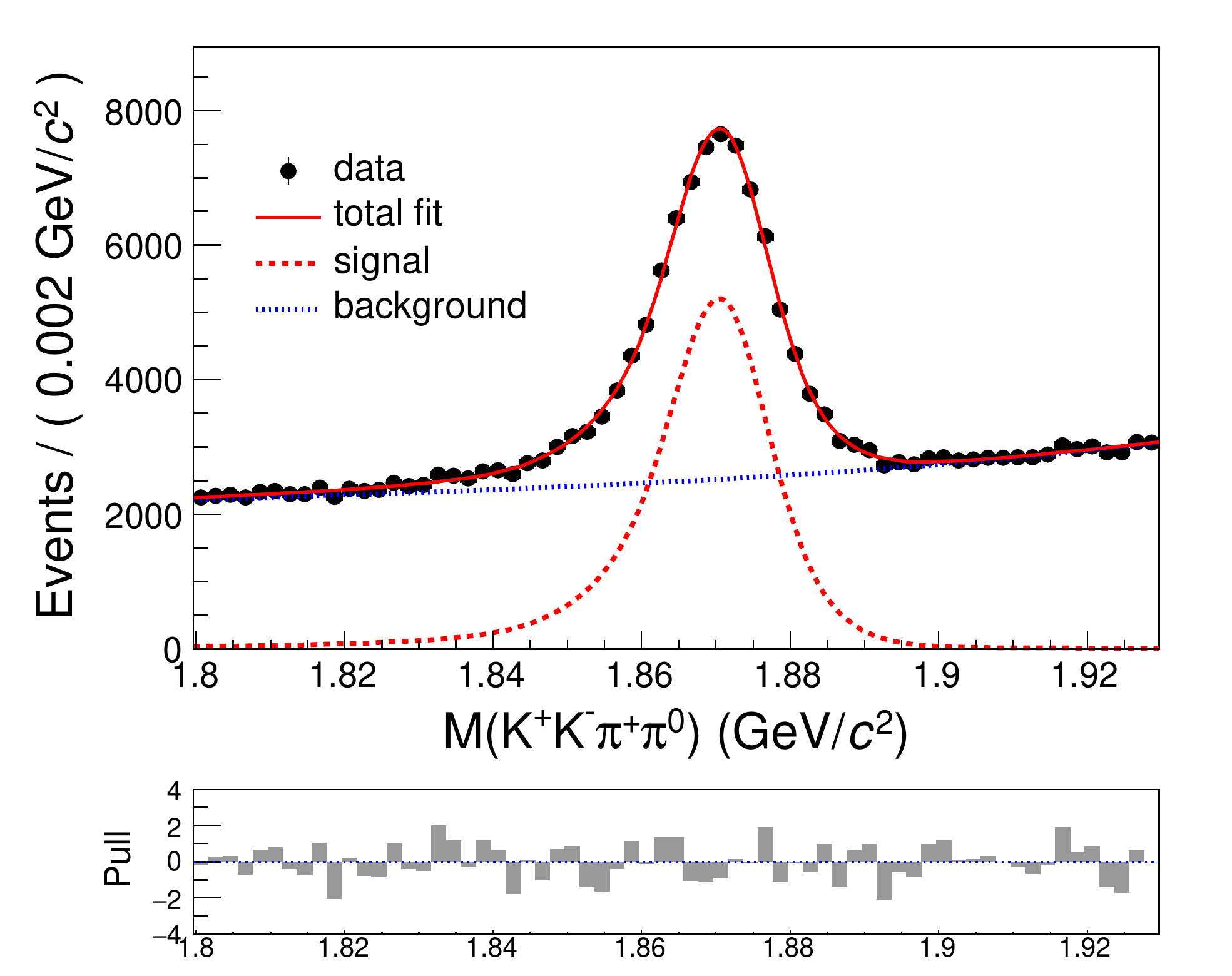}%
  \put(20,70){$\DpCS$}
  \end{overpic}\\
  \begin{overpic}[width=0.45\textwidth]{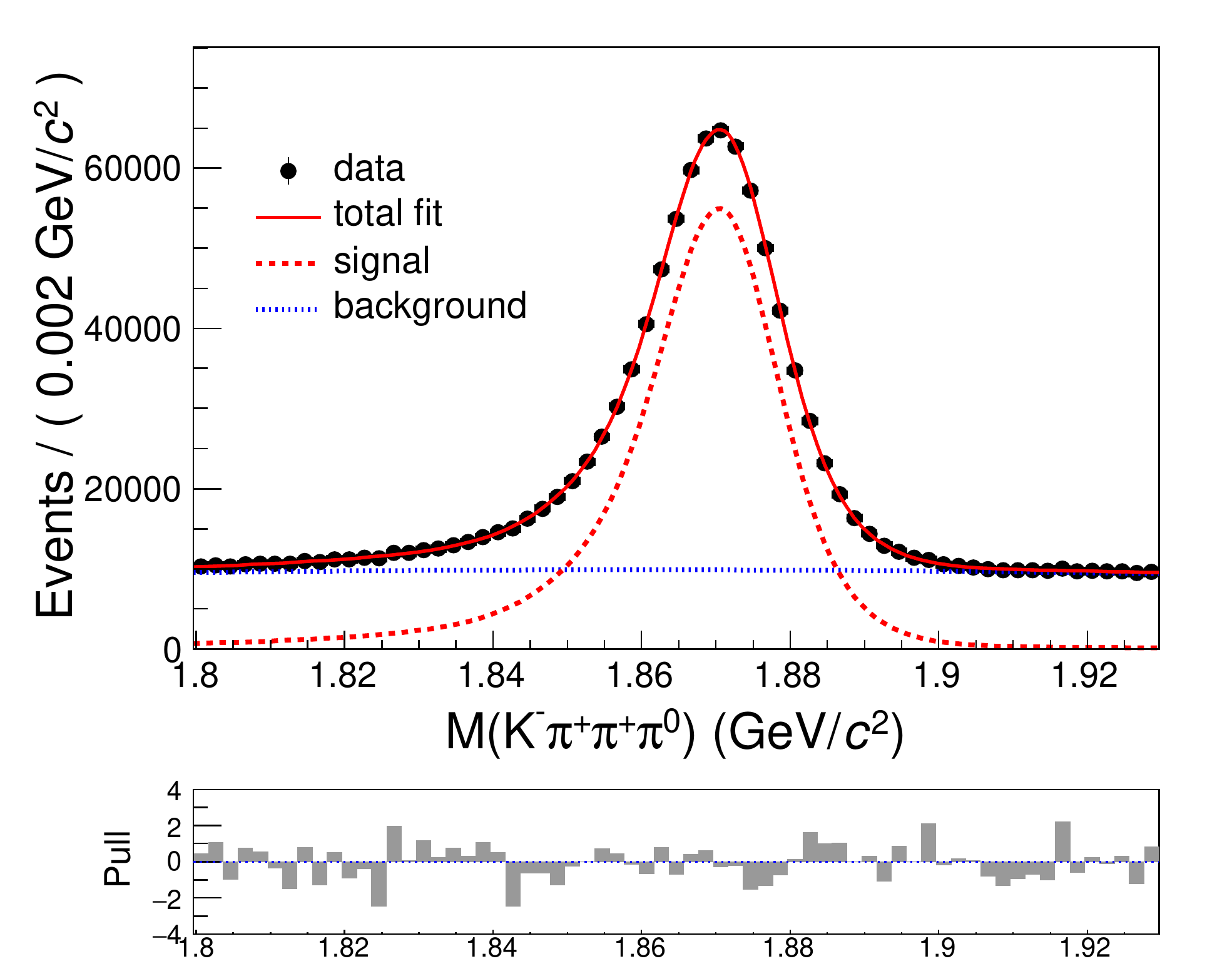}%
  \put(20,70){$\DpCF$}
  \end{overpic}%
    \vskip-10pt
  \caption{\label{fig:DpCS} Fit projections for $\DpCS$, and its normalization mode $\DpCF$. 
Data are plotted as filled circles with error bars.
The red solid, red dashed, and blue dashed curves denote the overall fit result, 
the signal component, and the background component, respectively.}
  \end{centering}
  \end{figure}

\begin{figure}[!htbp]
  \begin{centering}%
   \begin{overpic}[width=0.45\textwidth]{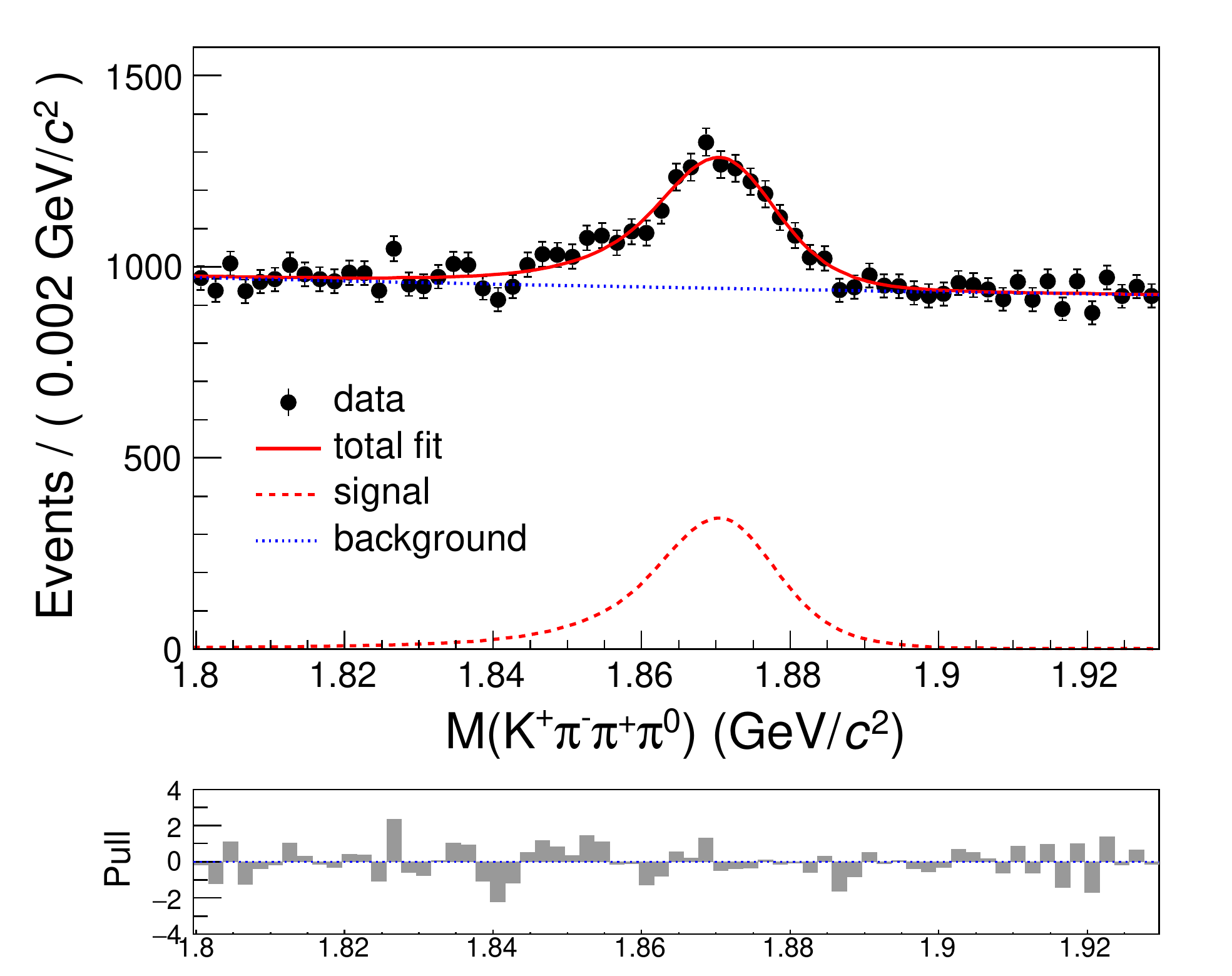}%
  \put(20,70){$\DpDCS$}
  \end{overpic}\\
  \begin{overpic}[width=0.45\textwidth]{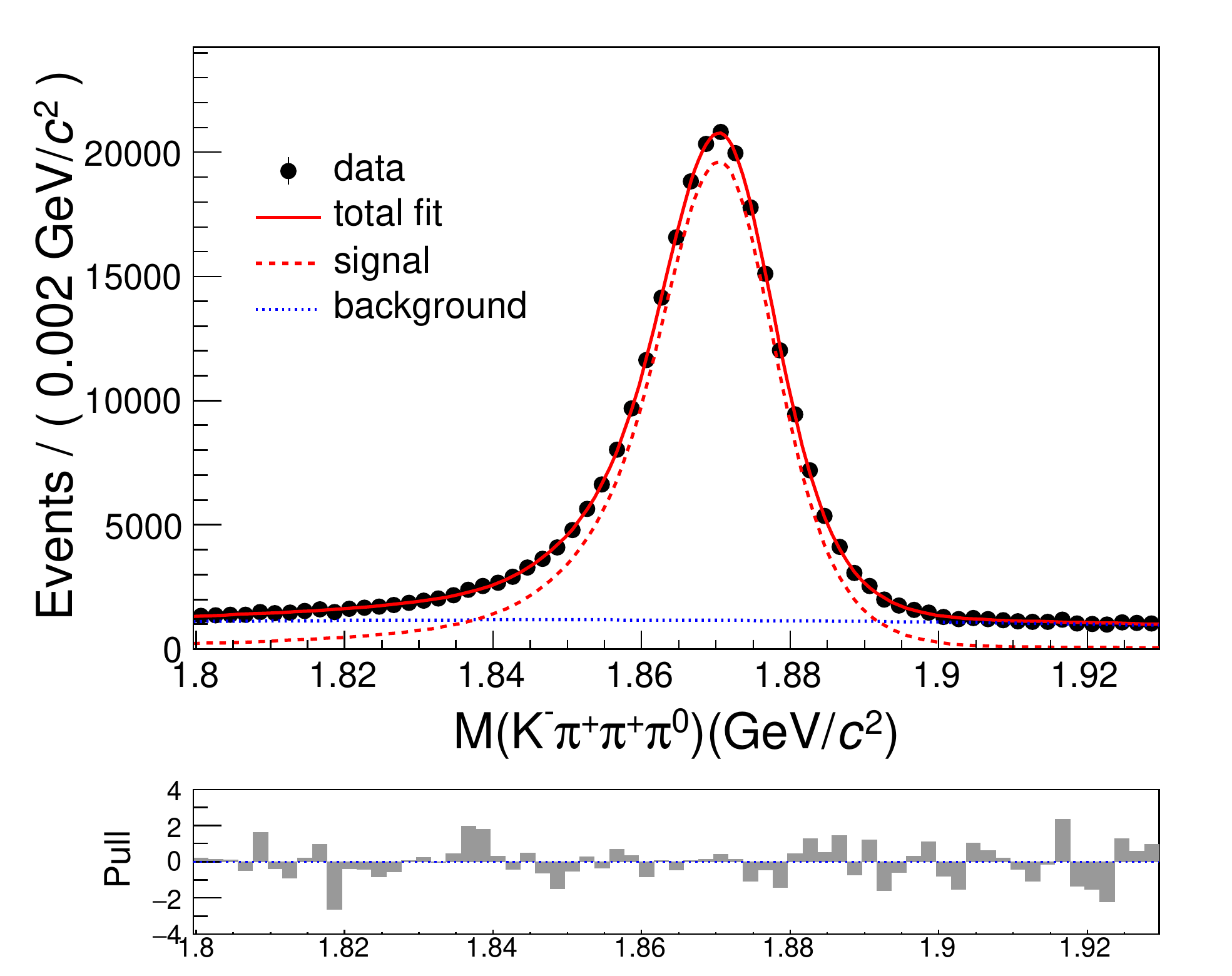}%
  \put(20,70){$\DpCF$}
  \end{overpic}%
  \vskip-10pt
  \caption{\label{fig:DpDCS} Fit projections for $\DpDCS$, and its normalization mode $\DpCF$. 
    The signal PDFs are the same for the two modes (see text).
    Data are plotted as filled circles with error bars.
    The red solid, red dashed, and blue dashed curves denote the overall fit result, 
    the signal component, and the background component, respectively.}
  \end{centering}
\end{figure}

\begin{figure}[!htbp]
  \begin{centering}%
  \begin{overpic}[width=0.45\textwidth]{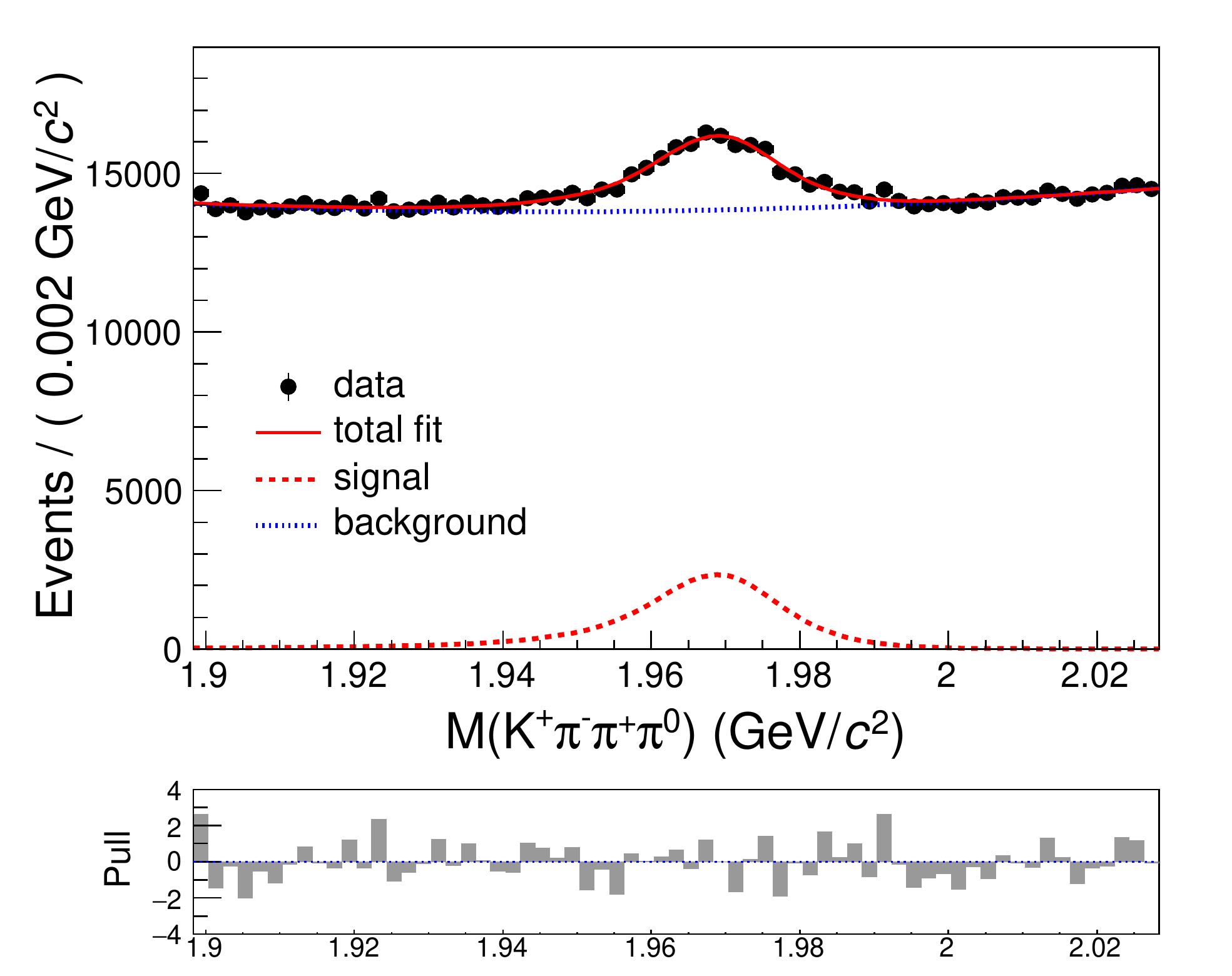} 
  \put(20,70){$\DsCS$}
  \end{overpic}\\
  \begin{overpic}[width=0.45\textwidth]{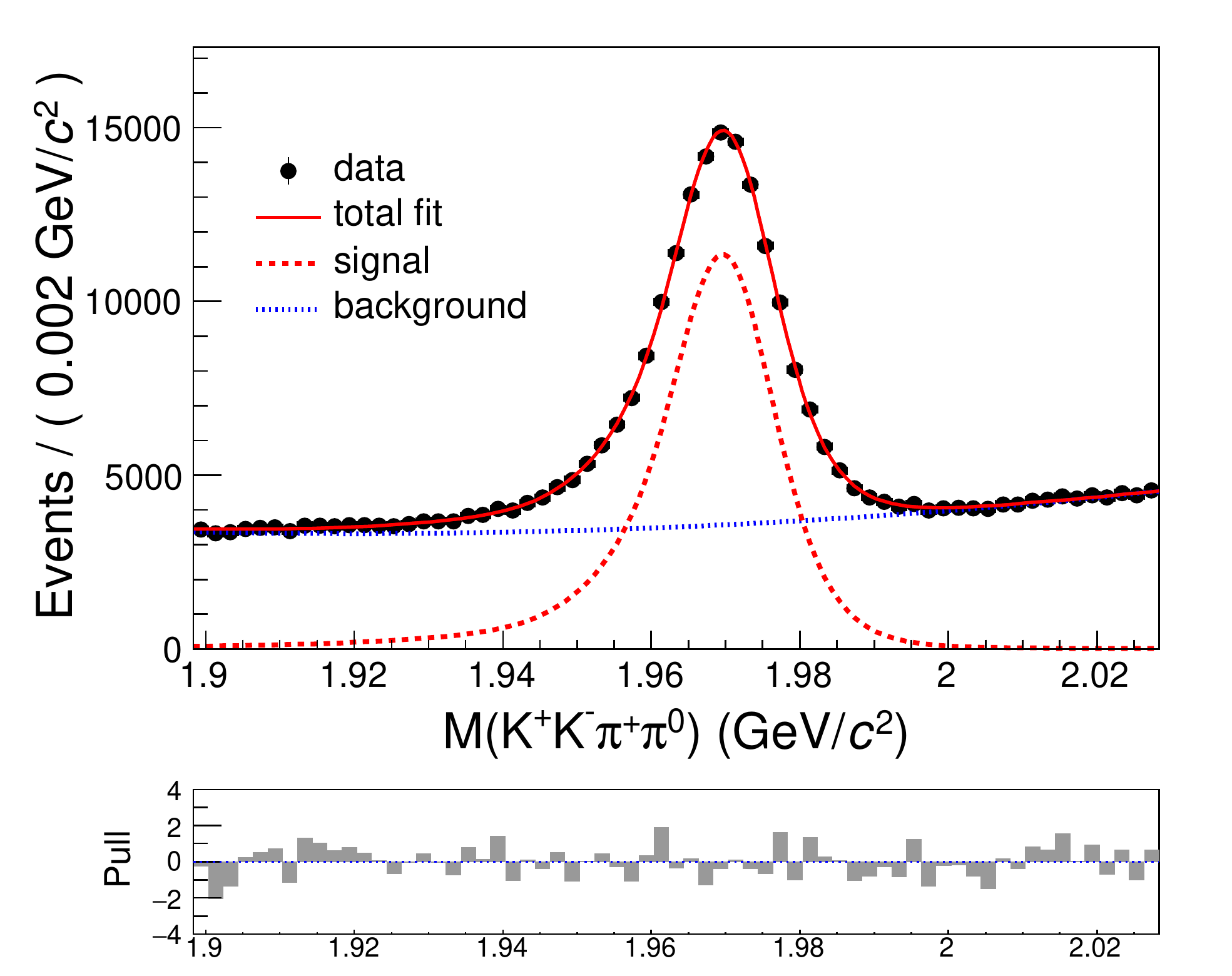} 
  \put(20,70){$\DsCF$}
  \end{overpic}
  \vskip-10pt
  \caption{\label{fig:DsCS} Fit projections for $\DsCS$, and its normalization mode $\DsCF$. 
  Data are plotted as filled circles with error bars.
The red solid, red dashed, and blue dashed curves denote the overall fit result, 
the signal component, and the background component, respectively.}
  \end{centering}
\end{figure}

Projections of the $M(D)$ fits 
are shown in Fig.~\ref{fig:DpCS} for $\DpCS$ 
and its normalization mode $\DpCF$; 
in Fig.~\ref{fig:DpDCS} for $\DpDCS$ and its 
normalization mode $\DpCF$; and in 
Fig.~\ref{fig:DsCS} for $\DsCS$ and its 
normalization mode $\DsCF$. 
Also plotted are the pulls, defined as 
$(N_{\rm data}-N_{\rm fit})/\sigma$, where $\sigma$ is the uncertainty on $N_{\rm data}$.
The pull distributions show that the fits describe the data satisfactorily.
The signal and background yields ($N_{\rm sig}$ and $N_{\rm bkg}$) 
obtained from the fits for the signal region, $\pm20$ MeV/$c^2$
around the nominal $\DpDsp$ mass, are listed in Table~\ref{tab:BRcal}.


The statistical significance of a signal yield 
is evaluated as the difference in the log likelihoods obtained from fits 
performed with and without a signal PDF.
For $\DpDCS$, we obtain $\Delta\ln\mathcal{L}=177$; 
as the number of degrees of freedom for the fit without a 
signal component is one less than that with a signal component,
this value corresponds to a statistical 
significance of greater than $10\sigma$.
For $\DsCS$, we obtain $\Delta\ln\mathcal{L}=594$.
In this case, the number of degrees of freedom 
without a signal component is three less than that with a signal component
(parameters $N_{\rm sig}$, $\mu$, and $\sigma_1$ are dropped),
and this value of $\Delta\ln\mathcal{L}$ corresponds to a statistical 
significance of greater than $10\sigma$. 
This measurement constitutes the first observation of this $\Dsp$ decay.

\begin{table*}[!htbp]
\setlength{\abovecaptionskip}{0cm}  
\setlength{\belowcaptionskip}{0.2cm} 
\begin{center}
\caption{\label{tab:BRcal}
Fitted signal yields ($N_{\rm sig}$) and background yields ($N_{\rm bkg}$) 
in the region $\pm 20~{\rm MeV}/c^2$ around the nominal $\DpDsp$ mass, 
and efficiency-corrected signal yields ($N_{\rm sig}^{\rm corr}$) for
(1) $\DpCS$; (2) $\DpDCS$; and (3) $\DsCS$. 
The normalization modes are $\DpCF$ for (1) and (2), and $\DsCF$ for (3). 
The $N_{\rm sig}$ values are not directly used to calculate 
$N_{\rm sig}^{\rm corr}$, which is calculated using Eq.~(\ref{eqn:correctedYields}). 
The normalization mode $\DpCF$ is used for both (1) and (2) but selected with different criteria in the two cases. Thus, the yields of this mode for (1) and (2) differ.
Also listed are ratios of branching fractions (see text).
The event yields are listed with their statistical uncertainty;
the branching fraction ratios are listed with both statistical and systematic uncertainties.
The right-most column lists the current world 
average~(WA)~\cite{bib:PRD102d052006,bib:PDG2021,bib:PRL125D141802,bib:arxiv2105D14310}.}
\begin{tabular}{ccccccc} \hline\hline
\multicolumn{2}{c}{Decay mode}	&  $N_{\rm sig}$  & $N_{\rm bkg}$	& $N_{\rm sig}^{\rm corr}(\times10^6)$    
& Branching fraction ratio   &  Current WA \\ \hline   
\multirow{2}{*}{(1)}  & $\DpCS$   & $49\,798\pm 564$    & $50\,463\pm 214$    & $2.352\pm 0.025$   
                          &  \multirow{2}{*}{$(11.32\pm 0.13\pm 0.26)\%$} & \multirow{2}{*}{$(10.6\pm 0.6)\%$} \\
 & $\DpCF$   & $602\,463\pm 1\,614$  & $197\,151\pm 570$    & $20.77\pm 0.07$      &    & \\ \hline 
\multirow{2}{*}{(2)}  & $\DpDCS$  & $3\,631\pm 198$     & $18\,879\pm 101$      & $0.303\pm 0.020$    
                & \multirow{2}{*}{$(1.68\pm 0.11\pm 0.03)\%$} & \multirow{2}{*}{$(1.89\pm 0.12)\%$} \\
 & $\DpCF$   & $208\,118\pm 707$   & $22\,327\pm 212$    & $18.06\pm 0.10$    &   & \\ \hline 
\multirow{2}{*}{(3)}  & $\DsCS$  & $26\,150\pm 1\,442$   & $277\,160\pm 582$   & $1.464\pm 0.052$   
                                         & \multirow{2}{*}{$(17.13\pm 0.62\pm 0.51)\%$} & \multirow{2}{*}{...} \\
& $\DsCF$   & $110\,261\pm 735$    & $71\,425\pm 263$    & $8.547\pm 0.059$       &   & \\
\hline \hline  
\end{tabular}
\end{center}  
\end{table*}

\section{Branching fractions}
\label{sec:BR}
To determine the branching fractions, we divide the signal 
yields by their respective reconstruction efficiencies. 
However, the reconstruction efficiency for a decay can 
vary across the
four-body phase space, and the distribution 
in phase space of these decays is unknown.
To reduce systematic uncertainty 
arising from the unknown decay distribution
(which often contains intermediate resonances), 
we correct the signal yields for reconstruction efficiencies 
in bins of phase space as follows.

\begin{figure*}[!hbtp]
\begin{centering}%
  \begin{overpic}[width=0.99\textwidth]{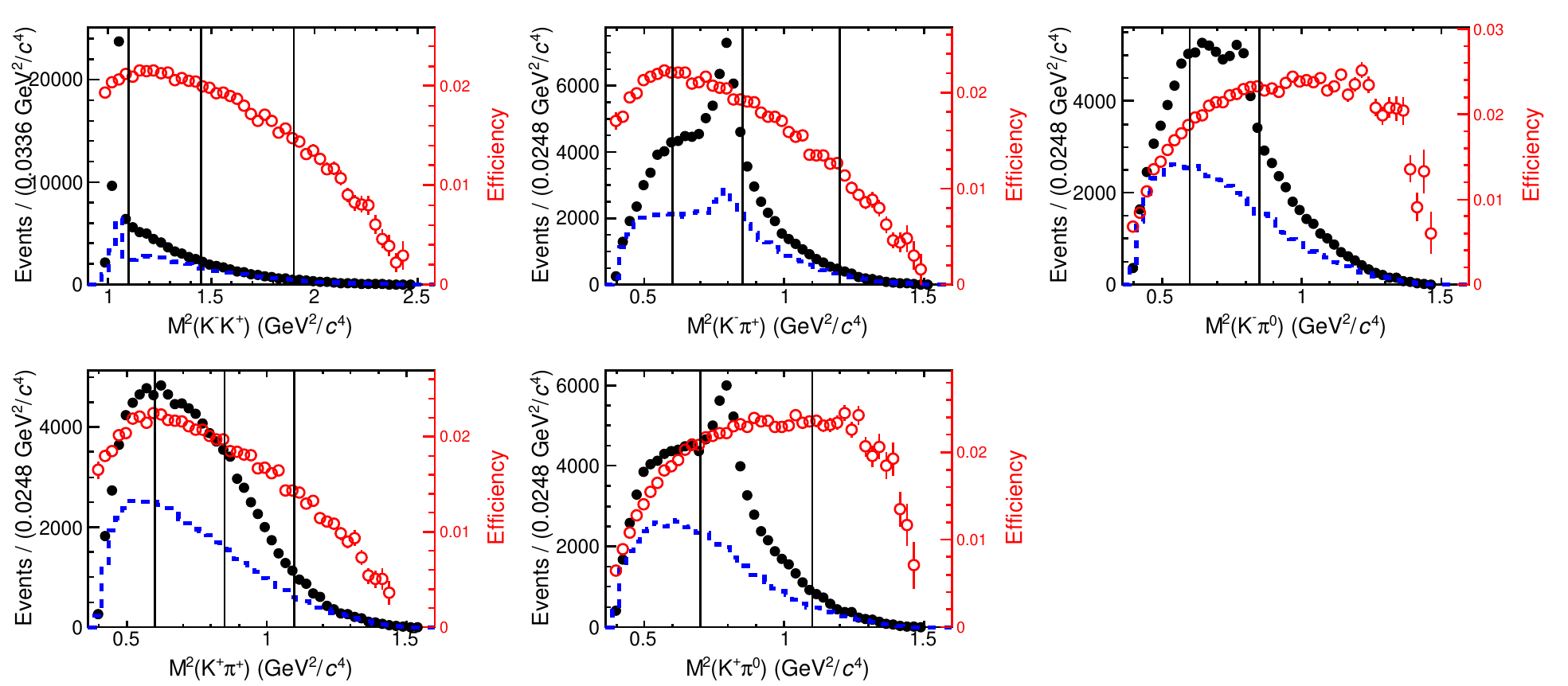}%
  \end{overpic}%
  \vskip-5pt   
\caption{\label{fig:DpCSeffVsDP}Invariant-mass-squared distributions for different combinations of final-state particles (i.e., projections of the five-dimensional distribution) for $\DpCS$ decays.
Events in the $M(D)$ signal region are plotted as 
filled black circles; events in the $M(D)$ sideband 
are plotted as blue dashed histograms; 
and signal efficiencies are plotted as red hollow circles. Black vertical lines denote the 
boundaries of the bins used for the efficiency correction.}
\end{centering}
\end{figure*}

\begin{figure*}[!hbtp]
\begin{centering}%
  \begin{overpic}[width=0.99\textwidth]{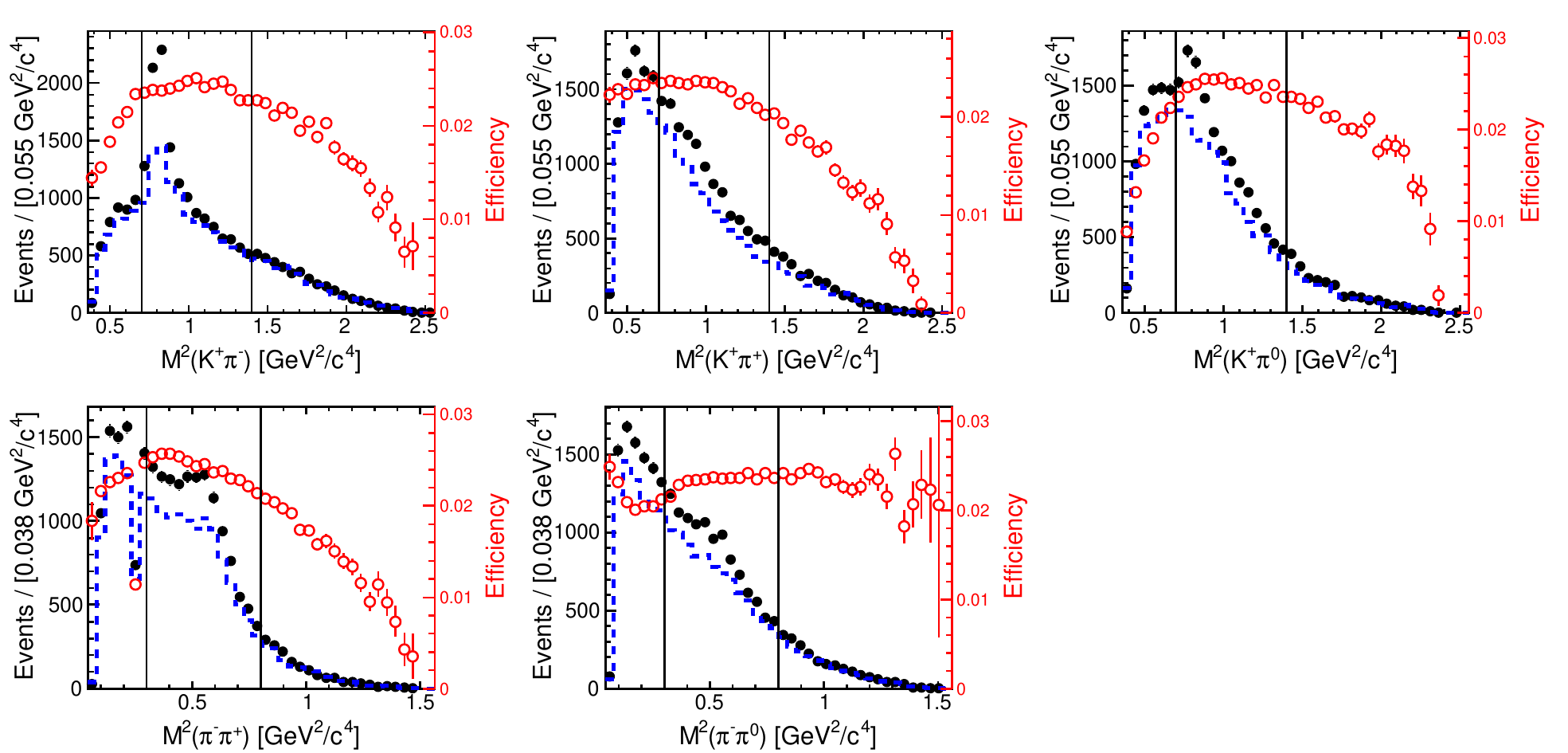}%
  \end{overpic}%
  \vskip-5pt   
\caption{\label{fig:DpDCSeffVsDP}
Invariant-mass-squared distributions for different combinations of final-state particles for $\DpDCS$ decays. Events in the $M(D)$ signal region are plotted as filled black circles; 
events in the $M(D)$ sideband are plotted as blue dashed histograms; and
signal efficiencies are plotted as red hollow circles. Black vertical lines denote the 
boundaries of the bins used for the efficiency correction.}
\end{centering}
\end{figure*}

\begin{figure*}[!hbtp]
\begin{centering}%
  \begin{overpic}[width=0.99\textwidth]{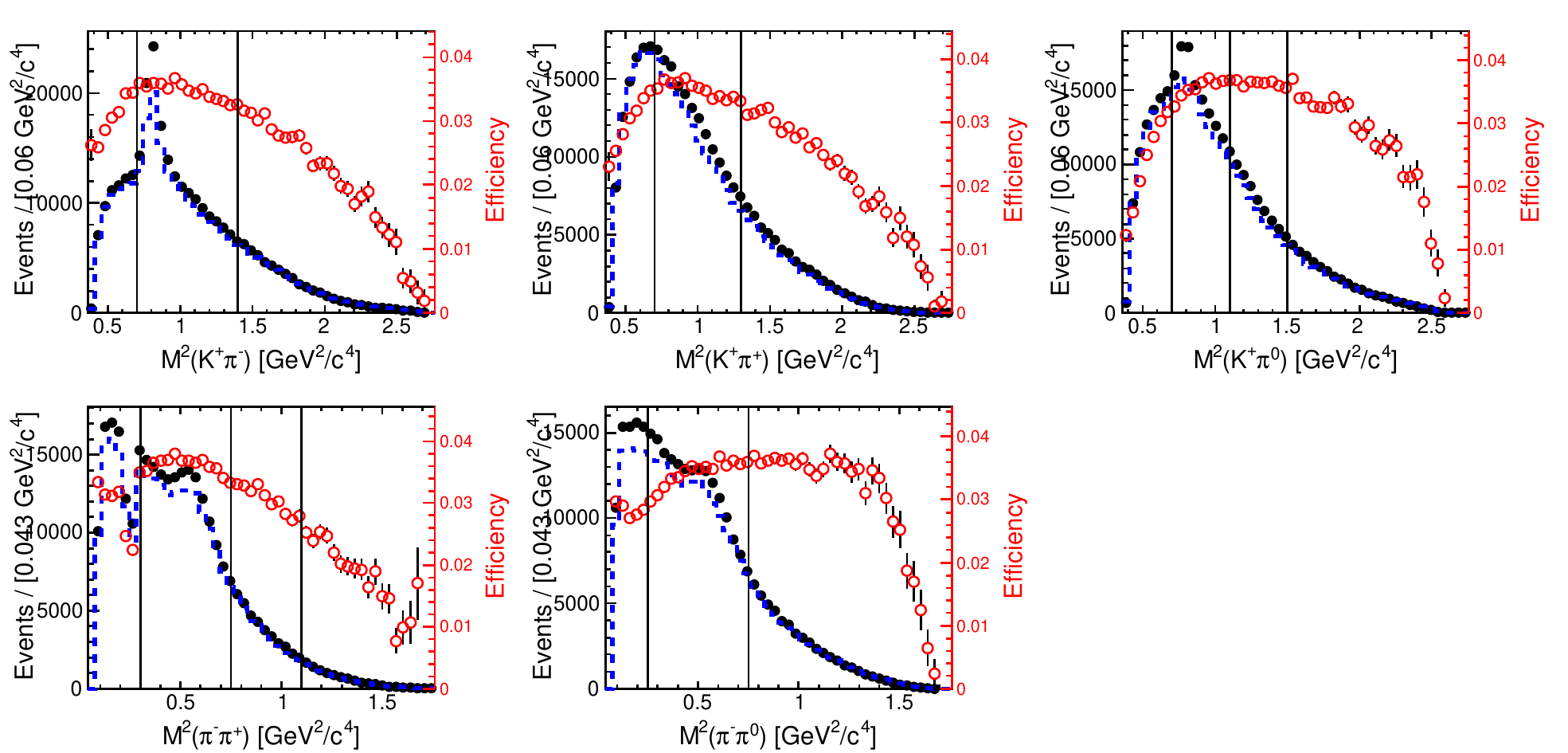}%
  \end{overpic}%
  \vskip-5pt   
\caption{\label{fig:DsCSeffVsDP}
Invariant-mass-squared distributions for different combinations of final-state particles for $\DsCS$ decays. Events in the $M(D)$ signal region are plotted as filled black circles; 
events in the $M(D)$ sideband are plotted as blue dashed histograms; and
signal efficiencies are plotted as red hollow circles. Black vertical lines denote the 
boundaries of the bins used for the efficiency correction.}
\end{centering}
\end{figure*}

For a $\DpDsp$ decay to four pseudoscalar particles in the final state, 
the phase space is five-dimensional~(5D). 
We thus correct the data for acceptance and reconstruction efficiency in bins of 5D phase space, where the bins are taken to be 
the invariant masses squared of five pairs of final-state particles~\footnote{This parameterization assumes that the detector acceptance and reconstruction efficiency are independent of the signs of $K$ and $\pi$ momenta in the $D$ rest frame (i.e., symmetric under a parity transformation). 
We have checked that any such dependence on the signs of momenta is negligible.}. 
These are calculated from fits subject to
the mass constraint $M(D)\!=\!m^{}_{\Dp}$ or $m^{}_{\Dsp}$.
The reconstruction efficiency is determined.
The efficiency-corrected signal yield is calculated as
\begin{eqnarray}
N_{\rm sig}^{\rm corr} & = & \sum_i \frac{ N_i^{\rm data} - N_{\rm bkg}\cdot f_i^{\rm bkg} }{ \varepsilon_i }\,, 
\label{eqn:correctedYields}
\end{eqnarray}
where 
$N_i^{\rm data}$, $f_i^{\rm bkg}$, and $\varepsilon_i$ are the
number of data events, the fraction of background events, and the 
reconstruction efficiency for bin~$i$. 
The summation runs over all bins.
The uncertainties on each term in Eq.~(\ref{eqn:correctedYields}), 
for each bin $i$, are propagated to obtain the overall uncertainty on $N^{\rm corr}_{\rm sig}$.
The bin sizes are chosen to minimize 
efficiency variations within the bins. 
There are 
576 bins for $\DpCS$ (i.e., 4$\times$4$\times$3$\times$4$\times$3); 
243 bins for $\DpDCS$; 
768 bins for $\DpCF$;
432 bins for $\DsCS$;
and 576 bins for $\DsCF$.
Invariant mass squared distributions for different combinations of final-state particles (i.e., projections of the five-dimensional distribution) are shown in 
Fig.~\ref{fig:DpCSeffVsDP} for $\DpCS$, 
Fig.~\ref{fig:DpDCSeffVsDP} for $\DpDCS$, 
and Fig.~\ref{fig:DsCSeffVsDP} for $\DsCS$.

The reconstruction efficiencies $\varepsilon_i$ are
determined from a large sample of MC events. These efficiencies 
include a correction for particle identification, to account for 
small differences observed between data and MC simulation. 
The correction, typically 0.93--1.03, is determined from 
a sample of $D^{*+}\to\Dz\pi^+,\,\Dz\to\Km\pip$ decays.
The fraction of background events
in the $i$th bin ($f_{i}^{\rm bkg}$) is obtained from 
the 5D distribution of events in the $M(D)$ sidebands
$-70~{\rm MeV}/c^2 < M(D)-m_{D} < -50~{\rm MeV}/c^2$ and 
$40~{\rm MeV}/c^2 < M(D)-m_{D} < 60~{\rm MeV}/c^2$.
An MC study shows that background in the signal region 
is well-described by background in the sidebands.
The background fractions must satisfy the 
constraint $\sum_{i}f^{\rm bkg}_{i}=1$. The efficiency-corrected signal yields 
obtained using Eq.~(\ref{eqn:correctedYields}) are listed in Table~\ref{tab:BRcal}.

The ratio of the efficiency-corrected yield for a CS mode to that for a CF mode
is equal to the ratio of branching fractions:
$N_{\rm sig}^{\rm corr}(\rm CS)/N_{\rm sig}^{\rm corr}(\rm CF) = \mathcal{B}(\rm CS)/\mathcal{B}(\rm CF)$.
Inserting values from Table~\ref{tab:BRcal}, we obtain
\begin{eqnarray}
\frac{\mathcal{B}(\DpCS)}{\mathcal{B}(\DpCF)} 	& = & (11.32 \pm 0.13 )\% \\
\frac{\mathcal{B}(\DpDCS)}{\mathcal{B}(\DpCF)} & = & (1.68 \pm 0.11 )\%   \\
\frac{\mathcal{B}(\DsCS)}{\mathcal{B}(\DsCF)} 	& = & (17.13 \pm 0.62 )\%\,,
\end{eqnarray}
where the uncertainties listed are statistical.

\section{Systematic uncertainties}\label{sec:syst}
Because we measure the ratio of branching fractions 
for decays with similar final states, most systematic
uncertainties cancel. 
The remaining uncertainties
are listed in Table~\ref{tab:BRsys} and evaluated as follows.

\begin{table}[tbh]
\setlength{\abovecaptionskip}{0cm}  
\setlength{\belowcaptionskip}{0.2cm} 
\begin{center}
\caption{\label{tab:BRsys} 
Fractional systematic uncertainties (in \%) for 
the following ratios of branching fractions: \\
(a) $\mathcal{B}(\DpCS)/\mathcal{B}(\DpCF)$;\\
(b) $\mathcal{B}(\DpDCS)/\mathcal{B}(\DpCF)$;
and \\
(c) $\mathcal{B}(\DsCS)/\mathcal{B}(\DsCF)$.}
\begin{tabular}{p{5.2cm} p{0.9cm} p{0.9cm} p{0.9cm}} \hline \hline 
Sources 	                &   (a) 	&   (b)     &   (c)     \\ \hline    	
PID efficiency correction 	&   $1.7$ 	&  ... 	    &   $1.7$ 	\\
Multiple-candidate selection&  $1.1$    &   $1.3$   &   $1.2$        \\ 
Signal parameterization		&   $0.5$ 	&   $0.5$   &   $1.1$ 	\\
$M(D)$ resolution           &   $0.5$       &   ...     &   $1.4$   \\
Binning                     &   $0.6$       &   $0.5$   &   $0.7$   \\
Background $M^2(p_ip_j)$ distribution 
                            &   $0.1$       &   $0.1$   &   $0.3$    \\
Efficiency correction bias  &   $0.6$       &   $0.8$   &   $1.1$    \\
\hline
Total uncertainty		&   $2.3$ 	&   $1.9$   &   $3.0$ 	 \\ 
\hline \hline 
\end{tabular}
\end{center}  
\end{table}

As a correction accounting for the difference in 
particle identification (PID)
efficiencies between data and MC is 
included in Eq.~(\ref{eqn:correctedYields}), we 
evaluate the uncertainty on this correction. 
Signal and normalization modes differ by 
the flavor of at most one track
in the final state. We evaluate the uncertainty introduced by this 
difference using a sample of ${D^{*+}\to D^0\pi^+}$, ${D^0\to K^{-}\pi^+}$ decays.
The resulting uncertainties are 
0.9\% for $\DpCS$ and 0.8\% for $\DpCF$; 
0.8\% for $\DsCS$ and 0.9\% for $\DsCF$. 
We thus assign 1.7\% as the systematic uncertainty for the
ratios of their branching fractions.

We consider uncertainty arising from the 
multiple-candidate selection procedure by 
keeping all candidates without best candidate selection. We refit the $M(D)$ distributions, redetermine the signal efficiency curves, and obtain the corrected yields with Eq.~(\ref{eqn:correctedYields}).
The resulting changes in the branching fractions are 
assigned as systematic uncertainties.

The uncertainty due to PDF parameters that are fixed in the fit for a 
signal yield is evaluated by sampling these parameters 
from a multivariate Gaussian distribution 
that accounts for their uncertainties and correlations,
and re-fitting for the signal yield. The procedure is 
repeated 1000 times,
and the root-mean-square of the distribution of fitted yields
is taken as the uncertainty due to the fixed parameters.

There are several sources of uncertainty in the efficiency correction procedure.
We first consider effects due to the $M(D)$ resolution. 
In Eq.~(\ref{eqn:correctedYields}), events satisfying $|M(D)-m_D|< 20~{\rm MeV}/c^2$ 
were used; however, data and MC 
could have different mass resolutions, and this would
bias the efficiency-corrected signal yield.
We evaluate the efficiency of the signal region requirement ($\eff_{\rm SR}$)
by integrating the signal PDFs over this region. The ratio of 
the efficiency for a signal mode to that of a normalization mode
is determined, both for data and MC. The ratio of 
these ratios
$[\eff_{\rm SR}^{\rm norm}/\eff_{\rm SR}^{\rm sig}\,({\rm data})] /
[\eff_{\rm SR}^{\rm norm}/\eff_{\rm SR}^{\rm sig}({\rm MC})]$ 
is calculated, and the difference from unity is taken as 
the systematic uncertainty due to the $M(D)$ resolution.
These uncertainties are 
$0.5\%$ for $\BR(\DpCS)/\BR(\DpCF)$ and $1.4\%$ for $\BR(\DsCS)/\BR(\DsCF)$. 
The uncertainty for  $\BR(\DpDCS)/\BR(\DpCF)$ is negligible, as the final 
states have the same particles.

We also consider uncertainty due to binning. 
We repeat the efficiency correction with 
a different number of bins, e.g., $4\!\times\!4\!\times\!4\!\times\!4\!\times\!4\!=\!1024$ bins for $\DpCS$,
and take the fractional change in the ratio of 
branching fractions as a systematic uncertainty.

We evaluate uncertainties arising from the 5D distribution of background events by applying 
a correction to the background distribution.
This correction, obtained from MC, is the ratio of the 
background distribution for events having $M(D)$ in the signal 
region to that for events having $M(D)$ in the sideband.
After applying this correction, the signal yield in
each efficiency bin is recalculated. The fractional 
change in the overall efficiency-corrected signal 
yield is assigned as a systematic uncertainty.

We check for bias in the efficiency correction due to possible intermediate 
resonances in the $\Dp$ or $\Dsp$ decay, also using MC simulation. 
The results for the efficiency-corrected signal yields are all 
consistent with input values; the small differences 
observed are conservatively assigned as systematic uncertainties.


The total systematic uncertainty is obtained by summing all
individual contributions in quadrature. The results are listed in Table~\ref{tab:BRsys}.

\section{Conclusion}
\label{sec:conclusion}

In summary, using 980~fb$^{-1}$ of data collected with the Belle detector, 
we observe the SCS decays $\DpCS$ and $\DsCS$, and the DCS decay $\DpDCS$.
The statistical significance of each mode is greater than~$10\sigma$. 
The branching fractions for these decays relative to the branching fractions 
for topologically similar CF decays are measured to be
\begin{eqnarray*}
\frac{\mathcal{B}(\DpCS)}{\mathcal{B}(\DpCF)} 	& = & (11.32 \pm 0.13 \pm 0.26 )\% \\
\frac{\mathcal{B}(\DpDCS)}{\mathcal{B}(\DpCF)} & = & (1.68 \pm 0.11 \pm 0.03)\% \\
\frac{\mathcal{B}(\DsCS)}{\mathcal{B}(\DsCF)} 	& = & (17.13 \pm 0.62 \pm 0.51 )\%\,, 
\end{eqnarray*}
where the uncertainties are statistical and systematic, respectively.
Taking $\sin\theta_C = 0.2257$~\cite{bib:PDG2021},
the second result above corresponds to 
$(5.83\pm0.42)\times\tan^4\theta_C$. 
This value is significantly larger than other measured ratios 
of DCS to CF branching fractions,
but it is consistent within $1.2\sigma$
with the large rate of $\DpDCS$ measured by
BESIII~\cite{bib:PRL125D141802,bib:arxiv2105D14310}.

Inserting world average values for the branching fractions of 
the normalization modes 
$\mathcal{B}(\DpCF)\!=\!(6.25\pm0.18)\%$~\cite{bib:PDG2021} and
$\mathcal{B}(\DsCF)\!=\!(5.51\pm0.19)\%$~\cite{bib:PDG2021,bib:DsToKKpipiz}, 
we obtain
\begin{eqnarray*}
\mathcal{B}(\DpCS) & = & \\
  & & \hskip-0.70in (7.08\pm 0.08 \pm 0.16\pm 0.20)\times 10^{-3} \\
\mathcal{B}(\DpDCS) & = & \\
  & & \hskip-0.70in (1.05\pm 0.07 \pm 0.02\pm 0.03)\times 10^{-3} \\
\mathcal{B}(\DsCS) & = & \\
  & & \hskip-0.70in (9.44\pm 0.34 \pm 0.28\pm 0.32)\times 10^{-3}\,,
\end{eqnarray*}
where the uncertainties are statistical, systematic, and from uncertainty in the branching fractions of the normalization modes, respectively.
The first two results are consistent with recent BESIII results~\cite{bib:PRD102d052006,bib:PRL125D141802,bib:arxiv2105D14310}
but have greater precision. The last result 
is the first measurement of this 
Cabibbo-suppressed decay.

\vskip5pt
\begin{acknowledgments}
This work, based on data collected using the Belle detector, which was
operated until June 2010, was supported by 
the Ministry of Education, Culture, Sports, Science, and
Technology (MEXT) of Japan, the Japan Society for the 
Promotion of Science (JSPS), and the Tau-Lepton Physics 
Research Center of Nagoya University; 
the Australian Research Council including grants
DP180102629, 
DP170102389, 
DP170102204, 
DE220100462, 
DP150103061, 
FT130100303; 
Austrian Federal Ministry of Education, Science and Research (FWF) and
FWF Austrian Science Fund No.~P~31361-N36;
the National Natural Science Foundation of China under Contracts
No.~11675166,  
No.~11705209;  
No.~11975076;  
No.~12135005;  
No.~12175041;  
No.~12161141008; 
Key Research Program of Frontier Sciences, Chinese Academy of Sciences (CAS), Grant No.~QYZDJ-SSW-SLH011; 
the Ministry of Education, Youth and Sports of the Czech
Republic under Contract No.~LTT17020;
the Czech Science Foundation Grant No. 22-18469S;
Horizon 2020 ERC Advanced Grant No.~884719 and ERC Starting Grant No.~947006 ``InterLeptons'' (European Union);
the Carl Zeiss Foundation, the Deutsche Forschungsgemeinschaft, the
Excellence Cluster Universe, and the VolkswagenStiftung;
the Department of Atomic Energy (Project Identification No. RTI 4002) and the Department of Science and Technology of India; 
the Istituto Nazionale di Fisica Nucleare of Italy; 
National Research Foundation (NRF) of Korea Grant
Nos.~2016R1\-D1A1B\-02012900, 2018R1\-A2B\-3003643,
2018R1\-A6A1A\-06024970, RS\-2022\-00197659,
2019R1\-I1A3A\-01058933, 2021R1\-A6A1A\-03043957,
2021R1\-F1A\-1060423, 2021R1\-F1A\-1064008, 2022R1\-A2C\-1003993;
Radiation Science Research Institute, Foreign Large-size Research Facility Application Supporting project, the Global Science Experimental Data Hub Center of the Korea Institute of Science and Technology Information and KREONET/GLORIAD;
the Polish Ministry of Science and Higher Education and 
the National Science Center;
the Ministry of Science and Higher Education of the Russian Federation, Agreement 14.W03.31.0026, 
and the HSE University Basic Research Program, Moscow; 
University of Tabuk research grants
S-1440-0321, S-0256-1438, and S-0280-1439 (Saudi Arabia);
the Slovenian Research Agency Grant Nos. J1-9124 and P1-0135;
Ikerbasque, Basque Foundation for Science, Spain;
the Swiss National Science Foundation; 
the Ministry of Education and the Ministry of Science and Technology of Taiwan;
and the United States Department of Energy and the National Science Foundation.
These acknowledgements are not to be interpreted as an endorsement of any
statement made by any of our institutes, funding agencies, governments, or
their representatives.
We thank the KEKB group for the excellent operation of the
accelerator; the KEK cryogenics group for the efficient
operation of the solenoid; and the KEK computer group and the Pacific Northwest National
Laboratory (PNNL) Environmental Molecular Sciences Laboratory (EMSL)
computing group for strong computing support; and the National
Institute of Informatics, and Science Information NETwork 6 (SINET6) for
valuable network support.

\end{acknowledgments}

\end{document}